\documentclass[12pt,manuscript]{aastex}

\newcommand{\kms}{\ensuremath{\rm km\,s^{-1}}}
\newcommand{\ms}{\ensuremath{\rm m\,s^{-1}}}
\newcommand{\gcmc}{\ensuremath{\rm g\,cm^{-3}}}

\newcommand{\teff}{\ensuremath{T_{\rm eff}}}
\newcommand{\logg}{\ensuremath{\log{g}}}
\newcommand{\vsini}{\ensuremath{v \sin{i}}}
\newcommand{\feh}{\rm{[Fe/H]}}

\newcommand{\rsun}{\ensuremath{R_\sun}}
\newcommand{\msun}{\ensuremath{M_\sun}}
\newcommand{\lsun}{\ensuremath{L_\sun}}

\newcommand{\rstar}{\ensuremath{R_\star}}
\newcommand{\mstar}{\ensuremath{M_\star}}
\newcommand{\loggstar}{\ensuremath{\logg_\star}}
\newcommand{\lstar}{\ensuremath{L_\star}}

\newcommand{\rpl}{\ensuremath{R_{\rm P}}}
\newcommand{\mpl}{\ensuremath{M_{\rm P}}}

\newcommand{\rhopl}{\ensuremath{\rho_{\rm P}}}
\newcommand{\loggpl}{\ensuremath{\logg_{\rm P}}}
\newcommand{\teq}{\ensuremath{T_{\rm eq}}}

\newcommand{\rjup}{\ensuremath{R_{\rm J}}}
\newcommand{\mjup}{\ensuremath{M_{\rm J}}}

\newcommand{\koicur}{Kepler-7}
\newcommand{\koicurb}{Kepler-7b}

\newcommand{\koicurCCra}{\ensuremath{19^{\mathrm{h}}14^{\mathrm{m}}19\fs{56}}}
\newcommand{\koicurCCdec}{\ensuremath{+41^{\circ}05'23\farcs{3}}}
\newcommand{\koicurCCkic}{KIC~5780885}
\newcommand{\koicurCCtwomass}{2MASS~19141956+4105233}
\newcommand{\koicurCCkicr}{12.815}			

\newcommand{\koicurLCar}{\ensuremath{7.22^{+0.16}_{-0.13}}}			
\newcommand{\koicurLCrprstar}{\ensuremath{0.08241^{+0.00030}_{-0.00043}}}	
\newcommand{\koicurLCimp}{\ensuremath{0.445^{+0.032}_{-0.044}}}			
\newcommand{\koicurLCi}{\ensuremath{86\fdg{5}\pm0.4}}				

\newcommand{\koicurLCP}{\ensuremath{4.885525\pm0.000040}}	
\newcommand{\koicurLCPshort}{4.886}				
\newcommand{\koicurLCPprec}{\ensuremath{4.885525}}		%
\newcommand{\koicurLCT}{\ensuremath{2454967.27571\pm0.00014}}	

\newcommand{\koicurSMEteff}{\ensuremath{5933\pm44}}	
\newcommand{\koicurSMEfeh}{\ensuremath{+0.11\pm0.03}}	
\newcommand{\koicurSMElogg}{\ensuremath{3.98\pm0.10}}	
\newcommand{\koicurSMEvsin}{\ensuremath{4.2\pm0.5}}	
\newcommand{\koicurMOOGteff}{\ensuremath{6000\pm75}}	
\newcommand{\koicurMOOGteffshort}{\ensuremath{6000}}	
\newcommand{\koicurMOOGfeh}{\ensuremath{+0.13\pm0.07}}	
\newcommand{\koicurMOOGlogg}{\ensuremath{4.00\pm0.10}}	

\newcommand{\koicurYYm}{\ensuremath{1.347^{+0.072}_{-0.054}}}		%
\newcommand{\koicurYYmshort}{\ensuremath{1.35}}				%
\newcommand{\koicurYYmlong}{\ensuremath{1.347^{+0.072}_{-0.054}}}	%
\newcommand{\koicurYYr}{\ensuremath{1.843^{+0.048}_{-0.066}}}		%
\newcommand{\koicurYYrshort}{\ensuremath{1.84}}				%
\newcommand{\koicurYYrlong}{\ensuremath{1.843^{+0.048}_{-0.066}}}	%
\newcommand{\koicurYYlogg}{\ensuremath{4.030^{+0.018}_{-0.019}}}	%
\newcommand{\koicurYYlum}{\ensuremath{4.15^{+0.63}_{-0.54}}}		%
\newcommand{\koicurYYage}{\ensuremath{3.5\pm1.0}}			%

\newcommand{\koicurRVK}{\ensuremath{42.9\pm3.5}}			
\newcommand{\koicurRVgamma}{\ensuremath{0}}				
\newcommand{\koicurRVmean}{\ensuremath{+0.40\pm0.10}}			

\newcommand{\koicurPPlogg}{\ensuremath{2.691^{+0.038}_{-0.045}}}	%
\newcommand{\koicurPParel}{\ensuremath{0.06224^{+0.00109}_{-0.00084}}}	
\newcommand{\koicurPPrho}{\ensuremath{0.166^{+0.019}_{-0.020}}}		%
\newcommand{\koicurPPrhoshort}{\ensuremath{0.17}}			%

\newcommand{\koicurPPm}{\ensuremath{0.433^{+0.040}_{-0.041}}}		%
\newcommand{\koicurPPmshort}{\ensuremath{0.43}}				%
\newcommand{\koicurPPmlong}{\ensuremath{0.433^{+0.040}_{-0.041}}}	%
\newcommand{\koicurPPr}{\ensuremath{1.478^{+0.050}_{-0.051}}}		%
\newcommand{\koicurPPrshort}{\ensuremath{1.48}}				%
\newcommand{\koicurPPrlong}{\ensuremath{1.478^{+0.050}_{-0.051}}}	%
\newcommand{\koicurPPteq}{\ensuremath{1540\pm200}}			%

\shortauthors{Latham et al.}
\shorttitle{\koicurb}
\slugcomment{Version resubmitted --- 26 December 2009}

\begin{document}

\title{\koicurb: A Transiting Planet with Unusually Low
Density\altaffilmark{\dagger}}

\altaffiltext{$\dagger$}
{Based in part on observations obtained at the W.~M.~Keck Observatory,
which is operated by the University of California and the California
Institute of Technology.}

\author{
David~W.~Latham\altaffilmark{1},
William~J.~Borucki\altaffilmark{2},
David~G.~Koch\altaffilmark{2},
Timothy~M.~Brown\altaffilmark{3},
Lars~A.~Buchhave\altaffilmark{1,4},
Gibor~Basri\altaffilmark{5},
Natalie~M.~Batalha\altaffilmark{6},
Douglas~A.~Caldwell\altaffilmark{7},
William~D.~Cochran\altaffilmark{8},
Edward~W.~Dunham\altaffilmark{9},
Gabor~F\H{u}r\'{e}sz\altaffilmark{1},
Thomas~N.~Gautier III\altaffilmark{10},
John~C.~Geary\altaffilmark{1},
Ronald~L.~Gilliland\altaffilmark{11},
Steve~B.~Howell\altaffilmark{12},
Jon~M.~Jenkins\altaffilmark{7},
Jack~J.~Lissauer\altaffilmark{2},
Geoffrey~W.~Marcy\altaffilmark{5},
David~G.~Monet\altaffilmark{13},
Jason~F.~Rowe\altaffilmark{14,2},
Dimitar~D.~Sasselov\altaffilmark{1}
}

\altaffiltext{1}{Harvard-Smithsonian Center for Astrophysics,
60 Garden Street, Cambridge, MA 02138}
\altaffiltext{2}{NASA Ames Research Center, Moffett Field, CA 94035}
\altaffiltext{3}{Las Cumbres Observatory Global Telescope, Goleta, CA 93117}
\altaffiltext{4}{Niels Bohr Institute, Copenhagen University, DK-2100 Copenhagen, Denmark}
\altaffiltext{5}{University of California, Berkeley, Berkeley, CA 94720}
\altaffiltext{6}{San Jose State University, San Jose, CA 95192}
\altaffiltext{7}{SETI Institute, Mountain View, CA 94043}
\altaffiltext{8}{University of Texas, Austin, TX 78712}
\altaffiltext{9}{Lowell Observatory, Flagstaff, AZ 86001}
\altaffiltext{10}{Jet Propulsion Laboratory/California Institute of Technology, Pasadena, CA 91109}
\altaffiltext{11}{Space Telescope Science Institute, Baltimore, MD 21218}
\altaffiltext{12}{National Optical Astronomy Observatory, Tucson, AZ 85719}
\altaffiltext{13}{US Naval Observatory, Flagstaff Station, Flagstaff, AZ 86001}
\altaffiltext{14}{NASA Postdoctoral Program Fellow}

\begin{abstract}
We report the discovery and confirmation of \koicurb, a transiting
planet with unusually low density.  The mass is less than half that of
Jupiter, $\mpl = \koicurPPmshort\,\mjup$, but the radius is fifty
percent larger, $\rpl = \koicurPPrshort\,\rjup$.  The resulting
density, $\rhopl = \koicurPPrhoshort\,\gcmc$, is the second lowest
reported so far for an extrasolar planet.  The orbital period is
fairly long, $P = \koicurLCPshort$ days, and the host star is not much
hotter than the Sun, $\teff = \koicurMOOGteffshort$\,K.  However, it
is more massive and considerably larger than the sun, $\mstar =
\koicurYYmshort\,\msun$ and $\rstar = \koicurYYrshort\,\rsun$, and
must be near the end of its life on the Main Sequence.
\end{abstract}



\section{INTRODUCTION}

The final test of the {\em Kepler} photometer at the end of
commissioning was a run of 9.7 continuous days in science mode, to
evaluate the noise performance of the instrument.  The {\em Kepler
Input Catalog} (KIC) was used to select fifty thousand isolated
targets, all with magnitudes brighter than 13.8 in the {\em Kepler}
passband, and with no nearby companions that would contaminate the
photometry.  The preliminary light curves from this test run were
inspected by team members with great excitement, and a few dozen
obvious planet candidates were quickly identified and passed on to the
team responsible for ground-based follow-up observations.  \koicur\
was observed but was not identified among the sample of initial
candidates.

After a gap of 1.3 days, normal science observations began for a full
list of more than 150,000 planet-search targets and continued for 33.5
days until interrupted on 15 June 2009, followed by a data download
and roll of the spacecraft to the summer orientation.  By the middle
of July the preliminary light curves were available for inspection,
and dozens of additional candidates were identified and passed on to
the follow-up team. This time \koicur\ was included.  Along with the
other candidates, \koicur\ was scrutinized for evidence of
astrophysical false positives involving eclipsing binaries.  It
survived this stage of the follow up and was then observed
spectroscopically for very precise radial velocities using the
FIber-fed Echelle Spectrograph (FIES) on the Nordic Optical Telescope
(NOT) during a ten night run in early October.  These observations
yielded a spectroscopic orbit that confirmed that an unseen companion
with a planetary mass was responsible for the dips in the light curve
observed by {\em Kepler}.

The KIC used ground-based multi-band photometry to assign an effective
temperature and surface gravity of $\teff = 5944$ K and $\logg =
4.27$\,(cgs) to \koicur, corresponding to a late F or early G dwarf.
Stellar gravities in this part of the H-R Diagram are notoriously
difficult to determine from photometry alone, and one of the
conclusions of this paper is that the star is near the end of its Main
Sequence lifetime, with a radius that has expanded to $\rstar =
\koicurYYr\,\rsun$ and a surface gravity that has weakened to 
$\logg = \koicurYYlogg$\,(cgs).  In turn this implies an inflated
radius for the planet, resulting in an unusually low density of
$\rhopl = \koicurPPrhoshort\,\gcmc$.  This conclusion is hard to
avoid, because the relatively long duration of the transit, more than
5 hours from first to last contact, demands a low density and expanded
radius for the star.

\section{KEPLER PHOTOMETRY}

\begin{figure}
\plotone{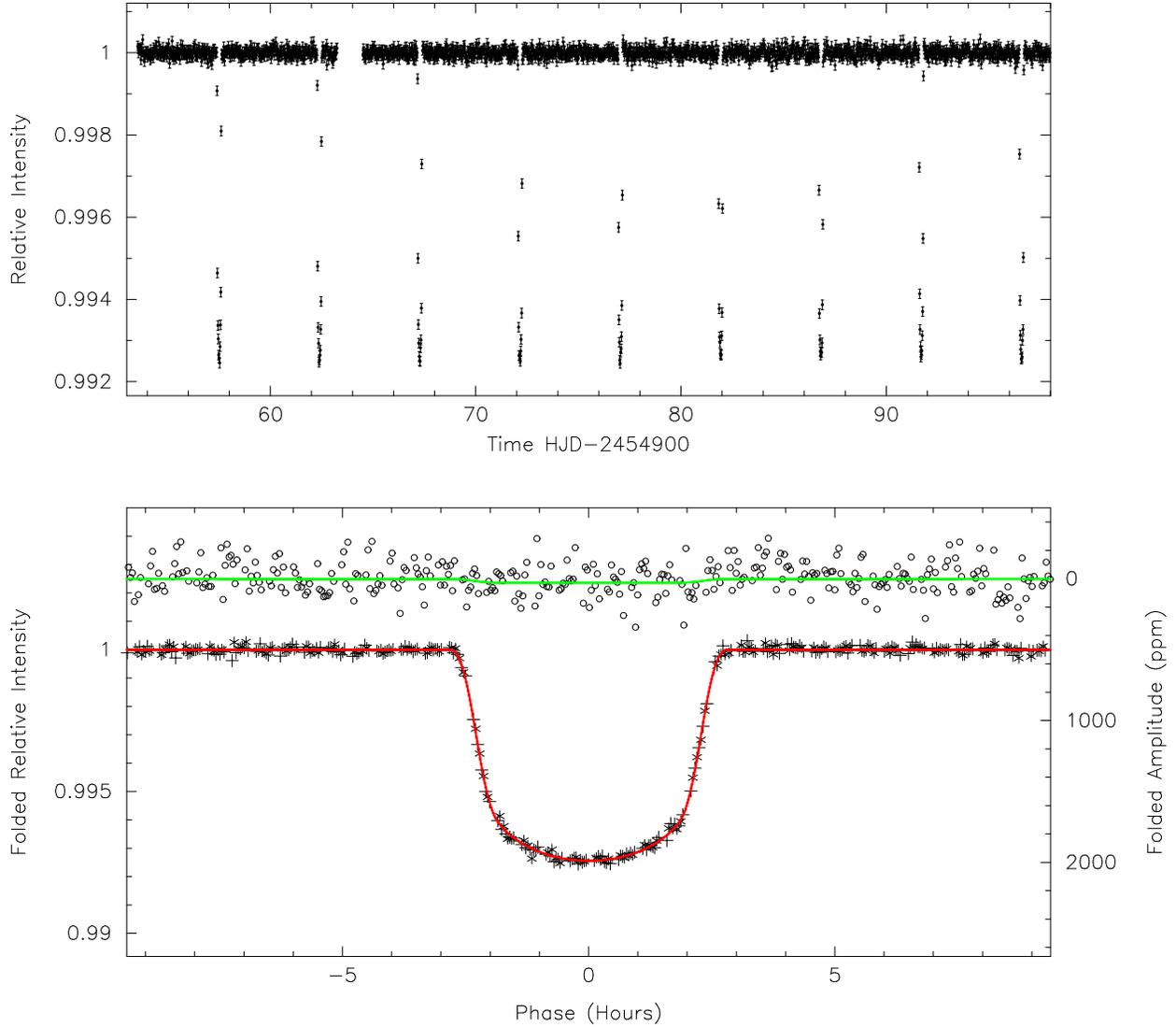}
\caption{
The detrended light curve for \koicur.  The time series for the entire
data set is plotted in the upper panel.  The lower panel shows the
photometry folded by the period $P = \koicurLCPprec$\,days.  The model
fit to the primary transit is plotted in red, and our attempt to fit a
corresponding secondary eclipse for a circular orbit is shown in green
with an expanded and offset scale.
\label{fig:lightcurve}}
\end{figure}

The light curve for \koicur\ (= \koicurCCkic, $\alpha = \koicurCCra,
\delta = \koicurCCdec$, J2000, KIC $r = \koicurCCkicr$\,mag) is
plotted in Figure~\ref{fig:lightcurve}.  The numerical data are
available electronically from the Multi Mission Archive at the Space
Telescope Science Institute (MAST) High Level Science Products (HLSP)
website\footnote{http://archive.stsci.edu/prepds/kepler\_hlsp}.  Only
a modest amount of detrending has been applied \citep{Koch:10} to this
time series of long cadence data (29.4-minute accumulations).  There
is no evidence for any systematic difference between alternating
events, which are plotted with $+$ and $\times$ symbols, supporting
the interpretation that all the events are primary transits.  Indeed,
there {\em is} weak evidence for a secondary eclipse centered at phase
0.5, as would be expected for a circular orbit, but the significance
is only about $2.4\sigma$. If this detection is real, it is not
inconsistent with the thermal emission expected from the planet for
reasonable assumptions \citep{Koch:10}.

\section{FOLLOW-UP OBSERVATIONS}

As described in more detail by \citet{Gautier:10}, the initial
follow-up observations of {\em Kepler} planet candidates involved
reconnaissance spectroscopy to look for evidence of a stellar
companion or a nearby eclipsing binary responsible for the observed
transits.  However, the follow-up team soon learned that the
astrometry derived from the {\em Kepler} images themselves, when combined
with high-resolution images of the target neighborhood, could provide
a very powerful tool for identifying background eclipsing binaries
blended with and contaminating the target images \citep{Batalha:10,
Monet:10}.  The astrometry of \koicur\ indicated a very slight image
centroid shift during transits of +0.1 millipixels in its CCD row direction
only.

The only star listed in the KIC that is closer than $30\arcsec$ to
\koicur\ and that can contribute significant light to the \koicur\
photometry is KIC~5780899, which is 4.4 mag fainter and lies at a
separation of $15.5\arcsec$.  KIC~5780899 cannot be the source of the
observed dips, because that would induce centroid shifts of about 25
millipixels.  If KIC~5780899 is constant and \koicur\ is the source of
the transits, the predicted shifts are in the right direction and have
an amplitude of roughly 0.1 millipixels if a quarter of KIC~5780899's
light leaks into the \koicur\ aperture.  Thus KIC~5780899 provides a
satisfactory explanation for the observed shifts.

To check for very close companions, a speckle observation of \koicur\
was obtained by S.~Howell with the WIYN 3.5-m telescope on Kitt Peak.
It showed no companions in a $2\arcsec$ box centered on \koicur.
Subsequently, images obtained by H.~Isaacson with the HIRES guider on
Keck 1, and independently by G.~Mandushev with the 1.8-m Perkins
telescope and PRISM camera at the Lowell Observatory and by N.~Baliber
with the LCOGT Faulkes Telescope North on Haleakala, Maui, all
detected a companion at a separation of $1.8\arcsec$ (just outside the
WIYN speckle window) and about 4.4 mag fainter in the red.  This
companion cannot be the source of the observed centroid shifts.  If
it is the source of the dips in the light curve, the centroid shifts
would have to be larger than 1 millipixel, and in the wrong direction.
If it is constant, the shifts would be much too small to detect.
However, this companion does dilute the photometry of \koicur\ with a
contribution of about 2.1\%.  Adding in a quarter of the light from
the more distant companion gives a total dilution of about
$2.5\pm0.4\%$.  This dilution has been included in the analysis of the
light curve.

Reconnaissance spectra obtained by M.~Endl and W.~Cochran with the
coud\'e echelle spectrograph on the 2.7-m Harlan J.~Smith Telescope at
the McDonald Observatory showed that there was no significant velocity
variation at the level of 1 \kms, and therefore that an orbiting
stellar companion could not be responsible for the observed transits.
Furthermore, there was no sign of a composite spectrum or
contamination by the spectrum of an eclipsing binary.  The McDonald
spectra were classified by L.~Buchhave by finding the best match
between the observed spectra and a library of synthetic spectra
calculated by J.~Laird for an extensive grid of stellar models
\citep{Kurucz:92} using a line list developed by J.~Morse.  This
yielded $\teff = 6000\pm125$\,K, $\logg = 4.0\pm0.2$\,(cgs), and
$\vsini = 4\,\kms$, very close to the final values reported in
Table~\ref{tab:parameters}.

\section{FIES SPECTROSCOPY}

The FIbre-fed Echelle Spectrograph (FIES) on the 2.5-m Nordic Optical
Telescope (NOT) at La Palma, was not originally designed with very
precise radial velocities in mind.  In particular, the fiber feed does
not incorporate a scrambler, there is no attempt to control the
atmospheric pressure (e.g. by housing the optics in a vacuum
enclosure), and there is no correction of the images for atmospheric
dispersion.  However, the spectrograph does reside in its own
well-insulated room with active control of the temperature to a few
hundreths K, with the result that the optics are quite stable.
Furthermore, FIES has good throughput, partly because the seeing is
often excellent at the NOT site, and an automatic guider keeps the
image well centered on a fiber $1.3\arcsec$ in diameter.  These
advantages encouraged us to develop specialized observing procedures
and a new data reduction pipeline with the goal of measuring radial
velocities to better than $10\,\ms$ for the relatively faint planet
candidates identified by {\em Kepler}.

To establish a wavelength calibration that tracks slow drifts during a
long exposure, we adopted the strategy of obtaining strong exposures
of a Thorium-Argon hollow cathode lamp through the science fiber
immediately before and after each science exposure.  Long science
exposures are divided into three or more sub-exposures, to allow
detection of and correction for radiation events.  Contamination by
scattered moonlight can be a serious problem for very precise
velocities of faint targets.  FIES does not yet have a separate
fiber for monitoring the sky brightness, so care is needed to avoid
the moon, especially if there are thin clouds.

A new reduction and analysis pipeline optimized for measuring
precise radial velocities was developed by L.~Buchhave.  After
extraction of intensity- and wavelength-calibrated spectra, relative
velocities are derived for each echelle order by cross correlation
against a combined template created by shifting all the observed
spectra of the same star to a common velocity scale and co-adding
them.  The final velocity for each observation is the mean of the
results for the individual orders, weighted by the number of detected
photons but not by the velocity information content.  Orders with very
low signal levels and orders contaminated by telluric lines are not
used. The internal error of the mean is estimated from the scatter
over the orders.

We observed \koicur\ with FIES for an hour on each of ten consecutive
nights in October 2009.  On every night we observed a standard star,
HD~182488, soon before \koicur\, and also soon after on half the
nights.  HD~182488 is conveniently located close to the {\em Kepler}
field of view and is known from HIRES observations over several years
to be stable to better than $3\,\ms$, and thus was adopted as the
primary velocity standard by the follow-up team.  Our 15 velocities
for HD~182488 show an rms of $7\,\ms$, with a slow drift pattern with
an amplitude of several \ms.  Therefore we interpolated a correction
to our velocity zero point for each observation of \koicur\ by
assuming that HD~182488 should not vary.  One of the 10 observations
was obtained through clouds and clearly showed a distortion of the
correlation peak due to contamination by scattered moonlight for
several of the blue orders.  This observation was rejected.  The
results for the other 9 observations are reported in
Table~\ref{tab:rvs}, including the variations in the line bisectors
and errors.

\begin{figure}
\plotone{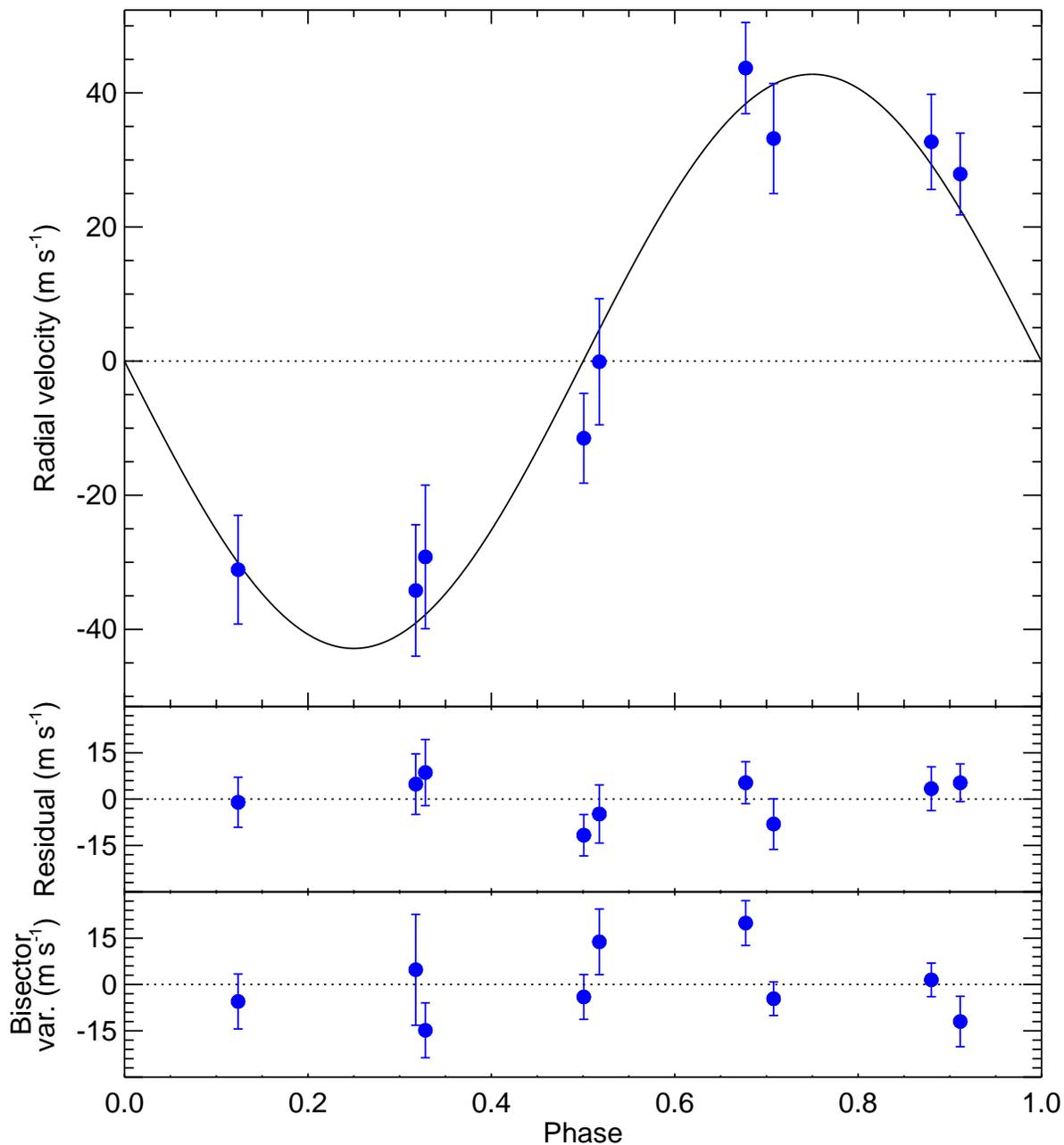}
\caption{
a) The orbital solution for \koicur. The observed radial velocities
obtained with FIES on the Nordic Optical Telescope are plotted
together with the velocity curve for a circular orbit with the period
and time of transit fixed by the photometric ephemeris.  The $\gamma$
velocity has been subtracted from the relative velocities here and in
Table~\ref{tab:rvs}, and thus the center-of-mass velocity for the orbital
solution is 0 by definition.
b) The velocity residuals from the orbital solution.  The rms of the
velocity residuals is 7.4\,\ms.
c) The variation in the bisector spans for the 9 FIES spectra.  The
mean value has been subtracted.
\label{fig:orbit}}
\end{figure}

We fit a circular orbit to the 9 velocities reported in
Table~\ref{tab:rvs}, adopting the photometric ephemeris, which leaves
the orbital semi-amplitude, $K$, and center-of-mass velocity,
$\gamma$, as the only free parameters.  A plot of this orbital
solution is shown in Figure~\ref{fig:orbit}, together with the
velocity residuals and the line bisector variations.  There is no
evidence of a correlation between the velocities and the bisectors,
which supports the interpretation that the velocity variations are due
to a planetary companion.  The orbital parameters are listed in
Table~\ref{tab:parameters}. Allowing the eccentricity to be a free
parameter reduced the velocity residuals by only a small amount and
yielded an eccentricity that was not significantly different from
circular.  A solution for a circular orbit using the velocities
uncorrected for the drifts exhibited by the standard star gave similar
velocity residuals, but a smaller value of $K$ by $7.6\,\ms$,
corresponding to an 18\% smaller mass.

The combined template spectrum for \koicur\ from FIES was analyzed by
A.~Sozzetti using MOOG\footnote{http://verdi.as.utexas.edu/moog.html},
to provide the stellar parameters needed to estimate the mass and
radius of the host star using stellar evolution tracks.  The critical
input parameters to the models are \teff\ and \feh, but the
spectroscopic \logg\ is also of interest for a consistency check.  A
spectrum of \koicur\ obtained by H.~Isaacson and G.~Marcy with HIRES
on Keck 1 was analyzed by D.~Fischer using SME \citep{Valenti:96},
with very similar results: $\teff = \koicurSMEteff$
vs.\ \koicurMOOGteff\ K, $\feh = \koicurSMEfeh$ vs.\ \koicurMOOGfeh, and
$\logg = \koicurSMElogg$ vs.\ \koicurMOOGlogg\,(cgs), for SME and MOOG,
respectively.  For the results reported in Table~\ref{tab:parameters},
we used the SME values.  The mean absolute velocity of \koicur,
$\koicurRVmean\,\kms$, was determined from the FIES observations by
adopting $-21.508\,\kms$ as the velocity for the standard star 
HD~182488.

\section{DISCUSSION\label{sec:discussion}}

The analysis of the {\em Kepler} photometry and the determination of the
stellar and planetary parameters for \koicur\ followed exactly the
procedures reported in \citet{Koch:10} and \citet{Borucki:10}.  The
results are reported in Table~\ref{tab:parameters}.  These results
were checked and confirmed by independent analyses carried out by
C.~Burke and G.~Torres.

The \koicur\ host star is not much hotter than the Sun, $\teff =
\koicurMOOGteff$\,K.  However, it is more massive and considerably
larger than the sun, $\mstar = \koicurYYm\,\msun$ and $\rstar =
\koicurYYrlong\,\rsun$, which puts it in a region of the H-R Diagram
near the end of its Main Sequence lifetime.  Indeed, the Yale-Yonsei
evolutionary tracks have hooks that cross at the position of \koicur,
and the probability distribution for the stellar mass has two peaks.
The stronger peak is for an evolutionary state not long before
Hydrogen burning in the core is exhausted with $\mstar = 1.362 \pm
0.040\,\msun$ and $\rstar = 1.857 \pm 0.047\,\rsun$, while the weaker
peak corresponds to a state soon after the star starts to evolve
rapidly, with $\mstar = 1.204 \pm 0.035\,\msun$ and $\rstar = 1.781
\pm 0.042\,\rsun$.  The mass for the evolved peak is 12\%
smaller, and the radius is 4\% smaller (as it must be to yield the
same stellar density).  The corresponding planetary radius is also 4\%
smaller, while the planetary mass is 8\% smaller (because of the
dependence on the 2/3 power of the system mass).  As our best guess
for the mass and radius of the host star and for the mass, radius, and
density of the planet, in Table~\ref{tab:parameters} we report the
mode and errors for the corresponding probability distributions.  This
takes into account all the possible evolutionary states for the host
star that are consistent with the observations.

\begin{figure}
\plotone{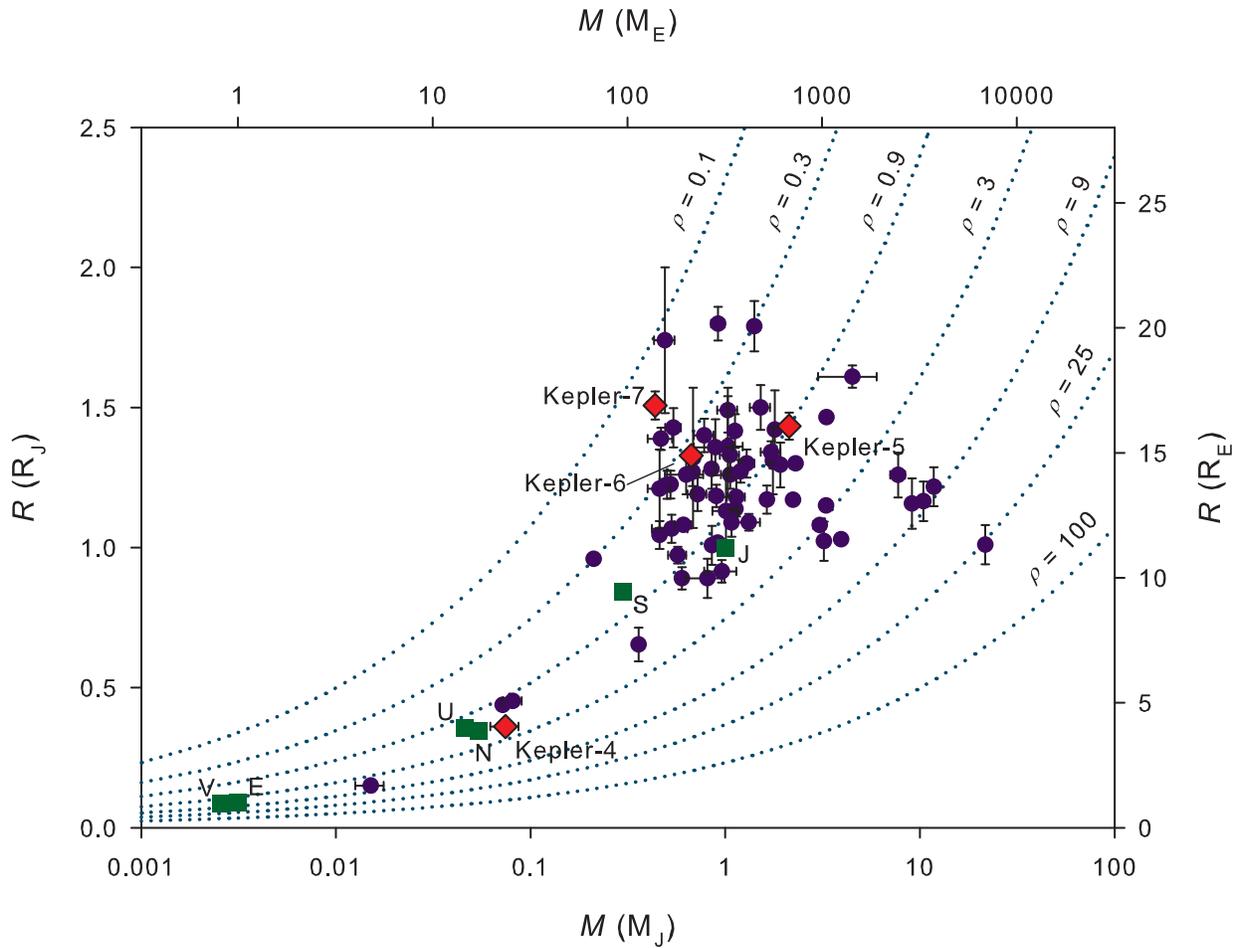}
\caption{
The Mass/Radius diagram for all the transiting planets with
known parameters as of 5 November 2009.  The four new {\em Kepler} planets are
labeled and plotted as diamonds.  \koicur\ has an unusually low density.
\label{fig:massradius}}
\end{figure}

The planetary radius is fifty percent larger than that of Jupiter,
$\rpl = \koicurPPrlong\,\rjup$, but the mass is less than half, $\mpl
= \koicurPPmlong\,\mjup$, which leads to an unusually low density of
$\rhopl = \koicurPPrho\,\gcmc$.  Among the known planets, only
WASP-17b appears to have a lower density \citep{Anderson:09}, although
the actual value for that planet is not yet well determined.  The
position of \koicurb\ on the mass/radius diagram is illustrated in
Figure~\ref{fig:massradius}, which plots all of the transiting planets
with known parameters as of 5 November 2009.  Because of possible
systematic errors in the radial velocities measured using FIES, the
mass of \koicurb\ may be smaller than we report by as much as 20\% or
even more.  However, the systematic error in the mass on the high side
is unlikely to be this large, because a larger orbital amplitude is
less vulnerable to systematic velocity errors.  For the planetary
radius, it is hard to avoid the conclusion that the planet is strongly
inflated, because the relatively long duration of the transit demands
a low density and expanded radius for the star.  A robust measure of
the transit duration is the time between the moment when the center of
the planet crosses the limb of the star during ingress and the
corresponding moment during egress.  A general formula for this
duration including the effect of orbital eccentricity is given by
\citet{Pal:10}, leading to a value of $4.63\pm0.06$ hours for \koicur.
We conclude that future observational refinements to the
characteristics of \koicurb\ are more likely to decrease the density
than increase it, with a significant uncertainty remaining as long as
the evolutionary state of the host star is uncertain.

\acknowledgements Many people have contributed to the success of the
{\em Kepler Mission}, and it is impossible to acknowledge them all by
name.  We offer our special thanks to the team of scientists and
programmers working with J.~M.~Jenkins to create the photometric
pipeline - H.~Chandrasekaran, S.~T.~Bryson, J.~Twicken, E~Quintana,
B.~Clarke, C.~Allen, J.~Li, P.~Tenenbaum, and H.~ Wu; to C.~J.~Burke
and G.~Torres for running independent checks of the analysis of the
\koicur\ light curve and system parameters; to J.~Andersen for help
with the FIES observations and unwavering moral support; to M.~Endl,
H.~Isaacson, D.~Ciardi, G.~Mandushev, N.~Baliber, and M.~Crane for
important contributions to the follow-up work; to A.~Sozzetti for his
analysis of the FIES combined template spectrum and to D.~Fischer for
her analysis of the HIRES template spectrum; to M.~Everett and
G.~Esquerdo for critical contributions to the KIC; to E.~Bachtel and
his team at Ball Aerospace for their work on the {\em Kepler}
photometer; to R.~Duren and R.~Thompson for key contributions to
engineering; and to C.~Botosh, M.~Haas, and J.~Fanson, for able
management.  DWL gratefully acknowledges partial support from NASA
Cooperative Agreement NCC2-1390 and the help of S.~Cahill and
L.~McArthur-Hines.  Funding for this Discovery mission is provided by
NASA's Science Mission Directorate.

{\it Facilities:} \facility{The Kepler Mission}, \facility{NOT (FIES)}, \facility{Keck:I (HIRES)},
\facility{WIYN (Speckle)}



\begin{deluxetable}{rrrrrr}
\tablewidth{0pc}
\tablecaption{Relative Radial-Velocity Measurements of \koicur{}\label{tab:rvs}}
\tablehead{
\colhead{HJD}                           &
\colhead{Phase}                         &
\colhead{RV}                            &
\colhead{\ensuremath{\sigma_{\rm RV}}}  &
\colhead{BS}                            &
\colhead{\ensuremath{\sigma_{\rm BS}}}  \\
\colhead{(days)}                        &
\colhead{(cycles)}                      &
\colhead{(\ms)}                         &
\colhead{(\ms)}                         &
\colhead{(\ms)}                         &
\colhead{(\ms)}
}
\startdata
2455107.37937 &  28.677 &$ +43.7 $&$\pm  6.8 $&$  +19.9 $&$\pm  7.3  $\\
2455108.36845 &  28.879 &$ +32.7 $&$\pm  7.1 $&$   +1.5 $&$\pm  5.4  $\\
2455110.50735 &  29.317 &$ -34.2 $&$\pm  9.8 $&$   +4.8 $&$\pm 17.9  $\\
2455111.40251 &  29.500 &$ -11.5 $&$\pm  6.7 $&$   -4.0 $&$\pm  7.2  $\\
2455112.41378 &  29.707 &$ +33.2 $&$\pm  8.2 $&$   -4.6 $&$\pm  5.4  $\\
2455113.40824 &  29.911 &$ +27.9 $&$\pm  6.1 $&$  -12.0 $&$\pm  8.2  $\\
2455114.44632 &  30.123 &$ -31.1 $&$\pm  8.1 $&$   -5.5 $&$\pm  8.9  $\\
2455115.44411 &  30.328 &$ -29.2 $&$\pm 10.7 $&$  -14.8 $&$\pm  8.9  $\\
2455116.37077 &  30.517 &$  -0.1 $&$\pm  9.4 $&$  +13.8 $&$\pm 10.6  $\\
\enddata
\end{deluxetable}


\begin{deluxetable}{lcc}
\tabletypesize{\scriptsize}
\tablewidth{0pc}
\tablecaption{System Parameters for \koicur \label{tab:parameters}}
\tablehead{\colhead{Parameter}	& 
\colhead{Value} 		& 
\colhead{Notes}}
\startdata
\sidehead{\em Transit and orbital parameters}
Orbital period $P$ (d)				& \koicurLCP		& A	\\
Midtransit time $E$ (HJD)			& \koicurLCT		& A	\\
Scaled semimajor axis $a/\rstar$		& \koicurLCar		& A	\\
Scaled planet radius \rpl/\rstar		& \koicurLCrprstar	& A	\\
Impact parameter $b \equiv a \cos{i}/\rstar$	& \koicurLCimp		& A	\\
Orbital inclination $i$ (deg)			& \koicurLCi 		& A	\\
Orbital semi-amplitude $K$ (\ms)		& \koicurRVK		& A,B	\\
Orbital eccentricity $e$			& 0 (adopted)		& A,B	\\
Center-of-mass velocity $\gamma$ (\ms)		& \koicurRVgamma	& A,B	\\
\sidehead{\em Observed stellar parameters}
Effective temperature \teff\ (K)		& \koicurSMEteff	& C 	\\
Spectroscopic gravity \logg\ (cgs)		& \koicurSMElogg	& C	\\
Metallicity \feh				& \koicurSMEfeh		& C	\\
Projected rotation \vsini\ (\kms)		& \koicurSMEvsin	& C	\\
Mean radial velocity (\kms)			& \koicurRVmean		& B	\\
\sidehead{\em Derived stellar parameters}
Mass \mstar (\msun)				& \koicurYYmlong	& C,D	\\
Radius \rstar (\rsun)  				& \koicurYYrlong	& C,D	\\
Surface gravity \loggstar\ (cgs)		& \koicurYYlogg		& C,D	\\
Luminosity \lstar\ (\lsun)			& \koicurYYlum		& C,D	\\
Age (Gyr)					& \koicurYYage		& C,D	\\
\sidehead{\em Planetary parameters}
Mass \mpl\ (\mjup)				& \koicurPPm		& A,B,C,D	\\
Radius \rpl\ (\rjup, equatorial)		& \koicurPPr		& A,B,C,D	\\
Density \rhopl\ (\gcmc)				& \koicurPPrho		& A,B,C,D	\\
Surface gravity \loggpl\ (cgs)			& \koicurPPlogg		& A,B,C,D	\\
Orbital semimajor axis $a$ (AU)			& \koicurPParel		& E	\\
Equilibrium temperature \teq\ (K)		& \koicurPPteq		& F
\enddata
\tablecomments{\\
A: Based on the photometry.\\
B: Based on the radial velocities.\\
C: Based on a MOOG analysis of the FIES spectra.\\
D: Based on the Yale-Yonsei stellar evolution tracks.\\
E: Based on Newton's version of Kepler's Third Law and total mass.\\
F: Assumes Bond albedo = 0.1 and complete redistribution.
}
\end{deluxetable}


\end{document}